\documentclass[12pt,preprint]{emulateapj}

\usepackage{natbib}

\shorttitle{Evidence of thermal conduction suppression}
\shortauthors{Wang et al.}

\begin{document}

\title{Evidence of thermal conduction suppression in a solar flaring loop by coronal seismology of slow-mode waves}

\author{Tongjiang Wang\altaffilmark{1,2}, Leon Ofman\altaffilmark{1,2,3},
Xudong Sun\altaffilmark{4}, Elena Provornikova\altaffilmark{1,2}, and Joseph M. Davila\altaffilmark{2}\\
Recevied: 22 Match 2015;~~~ Accepted: 28 August 2015}
\altaffiltext{1}{Department of Physics, Catholic University of America,
   620 Michigan Avenue NE, Washington, DC 20064, USA; tongjiang.wang@nasa.gov}
\altaffiltext{2}{NASA Goddard Space Flight Center, Code 671, Greenbelt, MD 20770, USA}
\altaffiltext{3}{Department of Geosciences, Tel Aviv University, Israel}
\altaffiltext{4}{W. W. Hansen Experimental Physics Laboratory, Stanford University, 
Stanford, CA 94305, USA}

\begin{abstract}
Analysis of a longitudinal wave event observed by the Atmospheric Imaging Assembly (AIA) 
onboard the {\it Solar Dynamics Observatory (SDO)} is presented. 
A time sequence of 131 \AA\ images reveals that a C-class flare occurred at 
one footpoint of a large loop and triggered an intensity disturbance (enhancement) propagating 
along it. The spatial features and temporal evolution suggest that a fundamental standing 
slow-mode wave could be set up quickly after meeting of two initial disturbances 
from the opposite footpoints. 
The oscillations have a period of $\sim$12 min and a decay time of $\sim$9 min. 
The measured phase speed of 500$\pm$50 km~s$^{-1}$ matches the sound speed
in the heated loop of $\sim$10 MK, confirming that the observed waves are of slow mode. 
We derive the time-dependent temperature and electron density wave signals
from six AIA extreme-ultraviolet (EUV) channels, and find that they are nearly in phase.
The measured polytropic index from the temperature and density perturbations is 
$1.64\pm0.08$ close to the adiabatic index of 5/3 for an ideal monatomic gas. The interpretation
based on a 1D linear MHD model suggests that the thermal conductivity is suppressed
by at least a factor of 3 in the hot flare loop at 9 MK and above. The viscosity 
coefficient is determined by coronal seismology from the observed wave
when only considering the compressive viscosity dissipation. We find that to interpret 
the rapid wave damping, the classical compressive viscosity coefficient needs 
to be enhanced by a factor of 15 as the upper limit. 

\end{abstract}


\section{Introduction}
Coronal seismology is a technique for measuring physical quantities of the corona 
by matching observations with magnetohydrodynamic (MHD) theory of waves in 
structured plasma \citep{rob83}. Considerable progress has been made in the study of
coronal seismology in the last decade \citep[see reviews by][]{nak05, dem12, liuw14}. 
From observations of propagating slow magnetoacoustic waves, \citet{van11} estimated 
the polytropic index to be $\sim$1 in the warm corona. The knowledge of the appropriate 
value of the polytropic index is important in hydrodynamic and MHD models of the solar 
and stellar coronae as well as of space plasmas \citep[e.g.][]{pud97, ril01, jac11}. 

Longitudinal hot loop oscillations were first discovered with the SUMER/$SOHO$ as periodic 
variations of Doppler shift in Fe\,{\sc xix} and Fe\,{\sc xxi} lines \citep{wan02, wan11}.
These oscillations were mainly interpreted as fundamental standing 
slow modes because their phase speed is close to sound speed in the loop, and there is a
quarter-period phase shift between the velocity and intensity oscillations 
in some observed cases \citep{wan03a, wan03b}. The initiation of the waves was often 
associated with small flares at the loop footpoint \citep{wan05}. 
3D MHD simulations show that the fundamental standing slow mode wave can be excited
by a velocity pulse or impulsive onset of flows near one footpoint of the loop 
\citep{sel09, ofm12}. These hot loop oscillations typically show a rapid decay.
MHD simulations by \citet{ofm02} suggested that thermal conduction is the dominant 
wave damping mechanism enhanced by nonlinear effect. The role of other physical effects
such as compressive viscosity, non-equilibrium ionization, shock dissipation, 
loop cooling, $etc$ was also studied theoretically 
\citep[see a review by][]{wan11, rud13, alg14}. 

Recently, \citet{kum13} reported a longitudinal wave event observed with the SDO/AIA,
showing similar physical properties as found previously in SUMER observations \citep{wan03b}.
However, the AIA wave was seen bouncing back and forth in the heated loop, suggesting 
that it may be a propagating mode in contrast to  the standing modes 
identified by SUMER. In this study we report the first SDO/AIA case
that shows clear signatures in agreement with a fundamental standing slow mode wave.
We find that the temperature and density perturbations in $\ga$9 MK plasma are nearly
in phase and the measured polytropic index accords well with the classical value (5/3) of the
adiabatic index for an ideal monoatomic gas, suggesting that the thermal conductivity 
in hot plasma is much weaker than predicted by the classical theory. This new finding 
may challenge our current understanding of thermal energy transport in solar and
stellar flares \citep{shi02}.

\begin{figure*}
\epsscale{1.0}
\plotone{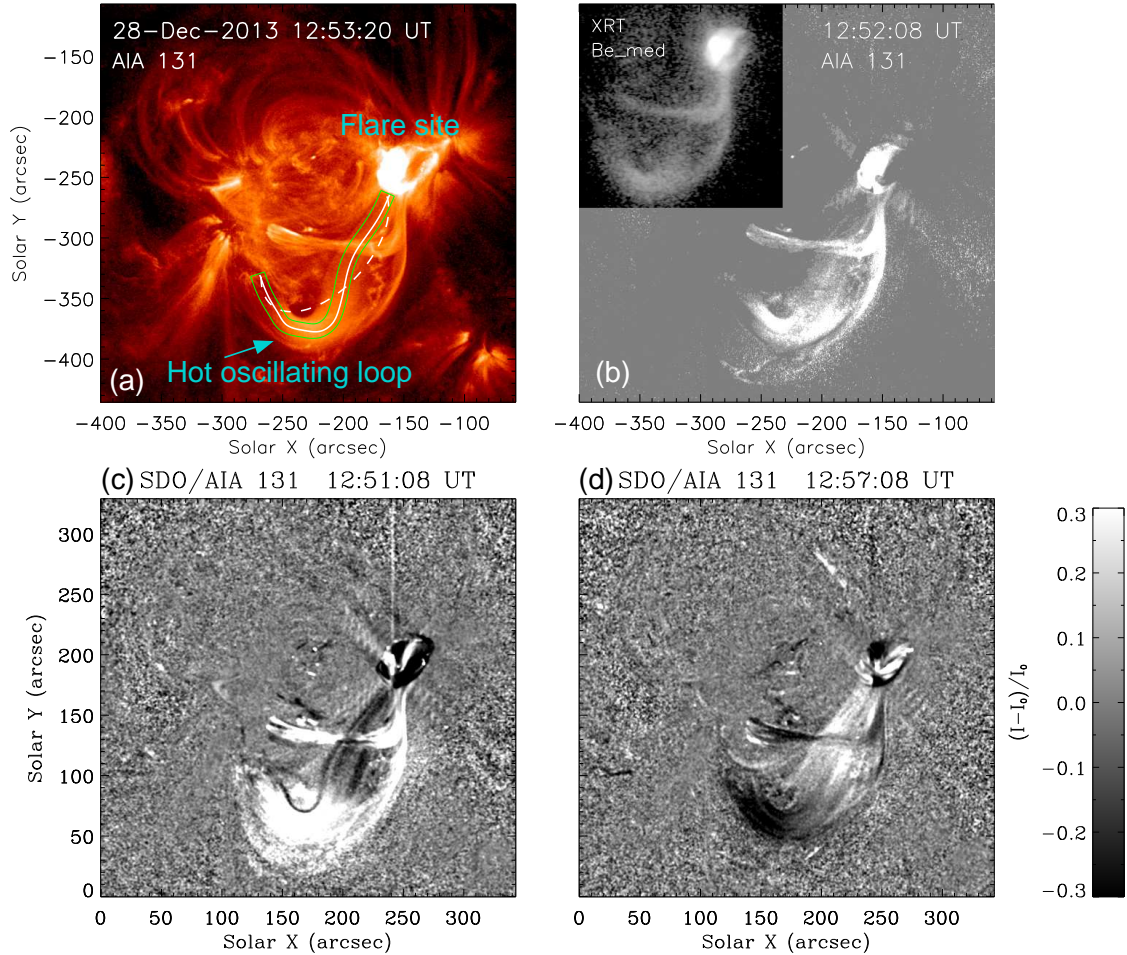}
\caption{ \label{fgflr} Observations of a longitudinal wave event  
on 2013 December 28. (a) AIA 131 \AA\ image. A sampled slice (green lines)
along the oscillating loop is used to construct the time-distance plot shown 
in Figure~\ref{fgosc}. The white thick curve indicates the reconstructed 3D loop
(same as the sampled track) and the dashed curve for a fitted circular model.
(b) AIA 131 \AA\ base-difference image relative to a pre-flare image at 12:40 UT. 
The inset shows a co-temporal Hinode/XRT image in the Be\_med filter. 
(c) and (d): The detrended 131 \AA\ images at 12:51 UT and 12:57 UT, 
showing the antiphase intensity perturbations at the opposite legs of the loop. 
(Animations are available in the online journal)}
\end{figure*}

\section{Observations and data analysis}
A GOES C3.0-class flare occurred on 2013 December 28 in NOAA Active Region (AR) 11936 
near disk center of the Sun. The flare began at 12:40 UT and peaked at 12:47 UT.
This flare excited longitudinal waves in a heated loop observed with $SDO$/AIA.
The AIA records the full Sun in ten extreme ultraviolet (EUV) and UV passbands 
with high spatio-temporal resolution and wide temperature coverage 
\citep[log$T$=3.7--7.3;][]{lem12}. We analyze the loop dynamics and thermal 
property using the AIA data.

The longitudinal waves were clearly detected in AIA 131 \AA\ (dominated by Fe\,{\sc xxi}, 
formed at $\sim$11 MK) band. The animation shows that a large loop that connected
the flare site at one end (marked with a curved cut in Figure~\ref{fgflr}(a)) developed 
quickly and exhibited intensity oscillations along the loop.  Similar emission features
seen in the AIA 131 \AA\ base-difference image and a co-temporal soft X-ray image
(Figure~\ref{fgflr}(b)) observed by Hinode/XRT \citep{gol07} indicate that 
the oscillating loop is hot. To emphasize the evolution of intensity perturbations, we processed
the 131 \AA\ images by subtracting the slowly-varying trend at each pixels, and then normalized
the perturbation to the trend, where the trend was derived using the Fourier low-pass 
($\ge$20 min) filtering. The detrended images show alternate intensity enhancements
at the opposite legs with rapid decay (Figures~\ref{fgflr}(c) and (d)). 
The animation shows that the loop oscillations lasted for about two periods 
before they faded out.

\begin{figure*}
\epsscale{1.0}
\plotone{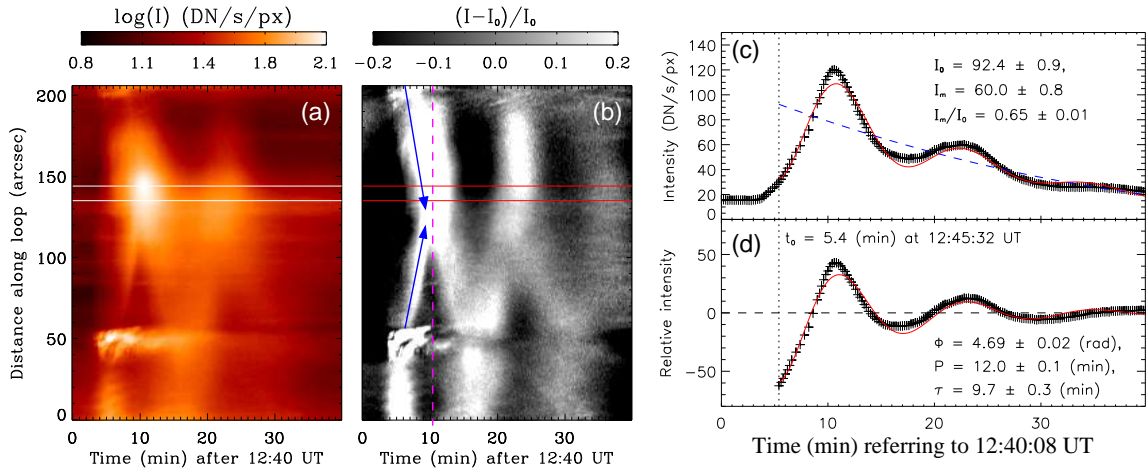}
\caption{ \label{fgosc} Measurements of the loop oscillations seen in 131 \AA.
(a) Time-distance plot for the sample wave path traced out in Figure~\ref{fgflr}(a), 
showing wave perturbations along the loop. In the plot, the distance is measured from the end 
near the flare site. (b) Same as (a) but with the slowly-varying trend subtracted 
at each position. The two arrows marked indicate two initial brightenings propagating 
in the opposite directions. (c) Time profile of the average counts extracted from 
a 9-arcsec wide region (narrow box in (a)). The solid curve shows a best fit which 
includes a parabolic trend (dashed line). (d) Detrended light curve with the best fit 
(solid line). }
\end{figure*}

\begin{figure*}
\epsscale{1.0}
\plotone{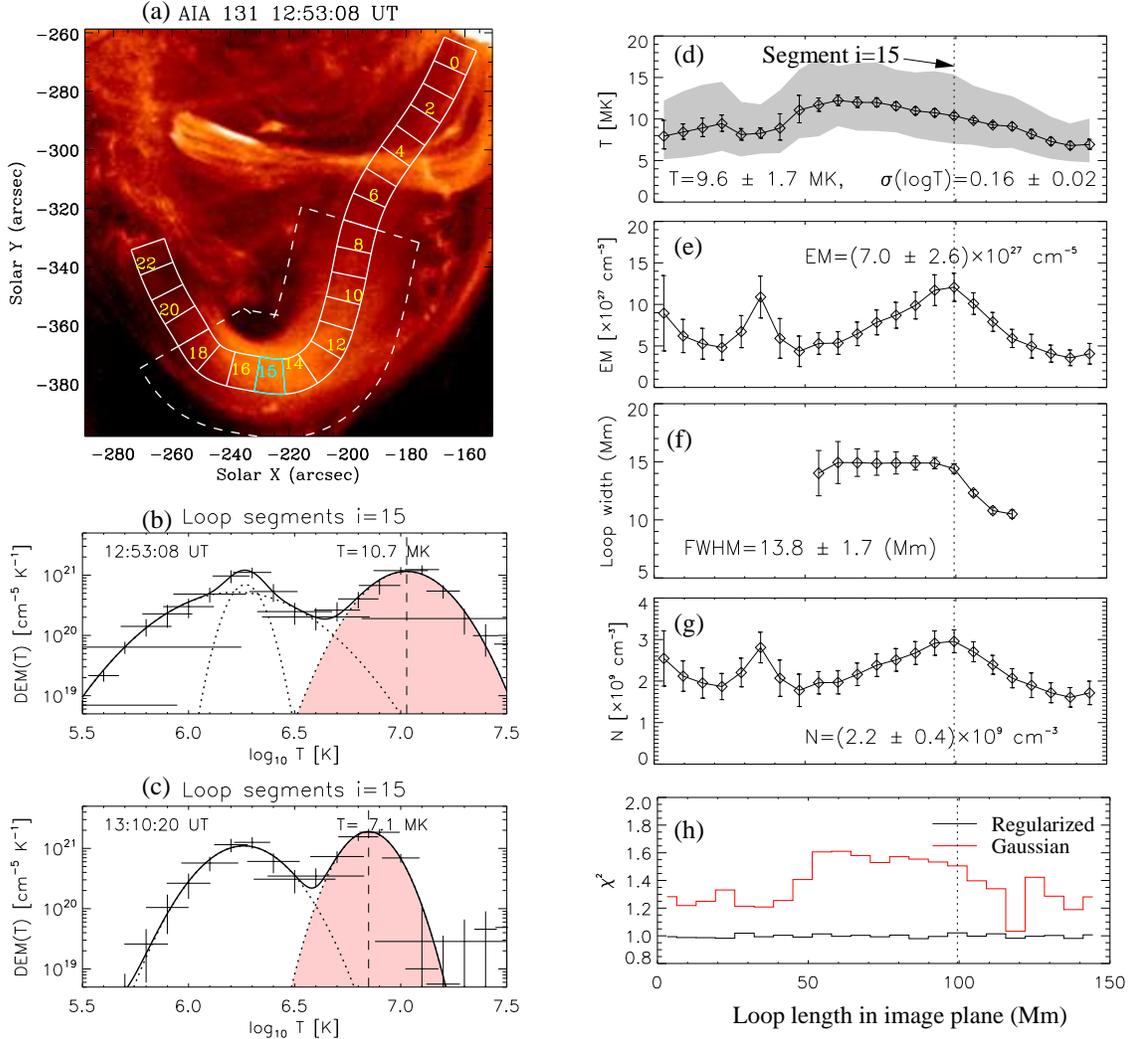}
\caption{ \label{fgdem} The DEM analysis of loop thermal properties. (a) AIA 131 \AA\ subimage 
of the analyzed loop. The traced loop is divided into 23 segments (solid lines) with 
segment $i$=0 near the flare site.  Each segment has a size of
15$\times$21 pixels. A 70-pixels wide region outlined with dashed lines is
selected for measuring the width of loop cross sections. The loop segment $i$=15 
is used for the analysis of EM and temperature evolution shown in Fig.~\ref{fgtnt}. 
Panels (b) and (c) show the regularized DEM inversion result 
for segment 15 at 12:53 UT and 17 min later. 
The DEM profile (crosses) is fitted to a triple-Gaussian model (curves) for isolating 
the foreground hot loop contribution (pink). (d) Loop temperature.
The gray band shows the temperature range ($\pm\sigma_T$, Gaussian width in Log$T$).
(e) Loop EM. (f) Loop width. (g) Electron density. 
(h) Reduced chi-squares ($\chi^2$) for the regularized DEM and the Gaussian model 
in data space. Error bars in (d)-(g) are the 1$\sigma$ fitting error.
(An animation is available in the online journal)}
\end{figure*}

To measure the wave properties we sampled the 131 \AA\ images from a 21-pixel (13 arcsec) 
wide slice (outlined in Figure~\ref{fgflr}(a)) over a period of 12:40--13:20 UT. 
We averaged the slices over the width and stacked them in time to construct 
a time-distance plot (Figure~\ref{fgosc}(a)). Figure~\ref{fgosc}(b) shows the corresponding 
detrended time-distance plot. The intensity oscillations along each leg are
nearly in-phase but anti-phase between the opposite legs matching the signature for 
a fundamental standing slow mode predicted by forward modelings \citep{tar08}.
Figure~\ref{fgosc}(c) shows the time profile of mean intensity 
for a cut selected at the leg with the brightest emission (at segment 15 shown in 
Figure~\ref{fgdem}(a)). We fitted the oscillatory signals with the function
\begin{equation}
I(t)=I_m {\rm sin}\left(\frac{2\pi(t-t_0)}{P}+\phi \right){\rm exp}\left(-\frac{t-t_0}{\tau}\right)
   +I_0(t), \label{eqoscft}
\end{equation}
where $I_m$, $P$, $\tau$, $\phi$, and $t_0$ are the amplitude, period, decay time, 
initial phase, and reference time, respectively, and $I_0(t)$ is a parabolic trend. 
The best fits to the light curve with and without the trend are shown in Figures~\ref{fgosc}(c) 
and \ref{fgosc}(d), respectively. 
The period of oscillation is $P=12.0\pm0.1$ min, the decay time $\tau=9.7\pm0.3$ min, 
and the initial amplitude relative to the trend is $I_m/I_0(t_0)=0.65\pm0.01$. 
 
The phase speed is important for identification of wave modes, which requires
an estimate of the loop length. We estimated the loop 3D geometry using the curvature 
radius maximization method which assumes the line-of-sight (LOS) coordinates of 
the observed loop to be the corresponding LOS coordinates of a circular model 
\citep{asc02, asc09}. Figure~\ref{fgflr}(a) shows the fitted circular model and 
the solution of the 3D geometry which has an identical 2D projection as observed.
We obtained the loop length $L$=179 Mm, about 33 Mm longer than that of its 2D projection.
The wave phase speed was estimated by $V_p$=2$L/P=500\pm50$ km~s$^{-1}$, where an uncertainty 
in $L$ is taken as $0.1L$. The phase speed is close to the sound speed of 480 km~s$^{-1}$ 
for the loop temperature $T\sim$10 MK, estimated using $c_s=152T_6^{1/2}$ km~s$^{-1}$, 
where $T_6=T$/[1 MK] and taking the adiabatic index $\gamma$=5/3.  
This result supports the interpretation of the observed waves as a fundamental slow mode. 

To conduct accurate diagnostics of the electron temperature and density of the oscillating loop 
we utilized a regularized differential emission measure (DEM) analysis on AIA images in six EUV
channels (94, 131, 171, 193, 211, and 335 \AA) \citep{han12, han13}. As an illustration, we
chose the images at $\sim$12:53 UT when the entire loop became obvious to trace, and divided
the entire loop into 23 subsections (Figure~\ref{fgdem}(a)). We performed the DEM 
inversion for each segment and fitted the DEM curve with a triple-Gaussian model to separate
a hot component ($\xi_{\rm hot}(T)$) from the background by assuming that the hot component 
comes from the foreground 
hot oscillating loop (Figures~\ref{fgdem}(b) and (h)). The application of this technique was 
detailed in \citet{sun13}. Figures~\ref{fgdem}(d) and (e) show the loop temperature
(the centroid of the hottest Gaussian component) and the loop emission measure 
($EM=\int\xi_{\rm hot}(T)dT$, the pink area in Figure~\ref{fgdem}(b)).
We measured the loop width by fitting the cross-sectional flux profile with a Gaussian function, 
and obtained the mean FWHM width, $w=13.8\pm1.7$ Mm for segments 8--18 (Figure~\ref{fgdem}(f)).
By using $n=\sqrt{EM/w}$ with a filling factor of unity (leading to a lower limit of $n$), 
the loop electron density was estimated (Figure~\ref{fgdem}(g)). Both the loop temperature 
and density are found to vary in a small range along the loop ($T$=7--12 MK and 
$n$=(1.6--3.0)$\times10^9$ cm$^{-3}$). 

\begin{figure*}
\epsscale{1.0}
\plotone{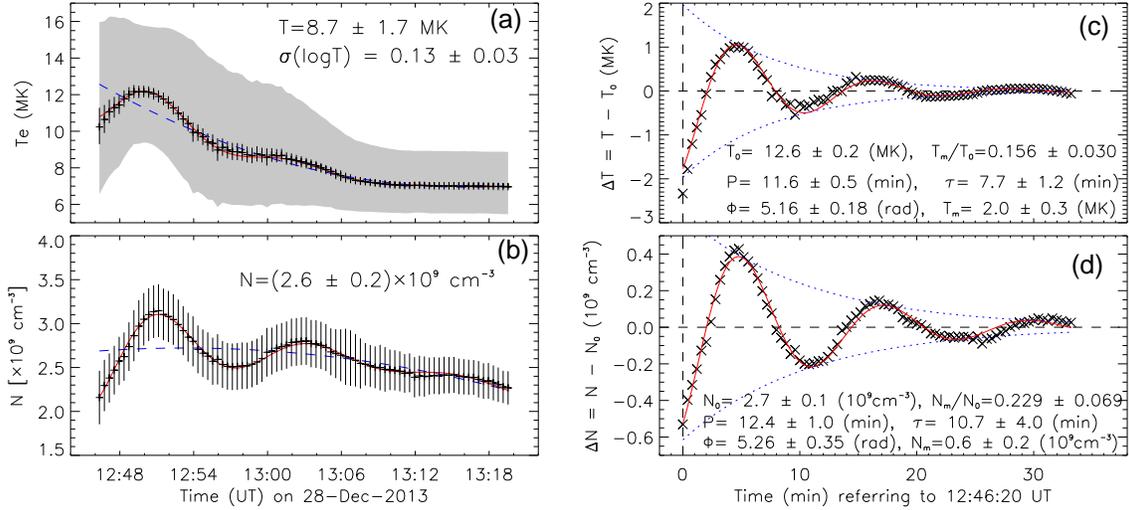}
\caption{ \label{fgtnt} Evolution of thermal properties for the loop segment 15. 
(a) Temperature. The red solid curve shows the best fit to 
the oscillatory signals, and the dashed curve is the parabolic trend. (b) Same as (a) but for
electron density.  The gray band and error bars in (a) and (b) have same meanings as 
in Figure~\ref{fgdem}. (c) Detrended time profile of the temperature (crosses) and the best
fit (red curve). (d) Same as (c) but for electron density.}
\end{figure*}

To study the time evolution of thermal properties, we focused on segment 15 with 
the maximum EM and used its average flux in six AIA bands to perform a time series of 
DEM inversions with a cadence of 24 s. By applying the triple-Gaussian DEM 
analysis (Figures~\ref{fgdem}(b) and (c) and the animation in the online version), 
we obtained time profiles of the loop temperature and electron density  
(Figures~\ref{fgtnt}(a) and (b)), where the loop width $w$ was taken as
a constant ($\sim$14 Mm) in calculation of the density evolution.
The mean temperature over the lifetime of oscillations
is found to be $T=8.7\pm1.7$ MK, and the mean density $n=(2.6\pm0.2)\times10^9$ cm$^{-3}$.
The wave signals in measured temperature and density are evident. We fitted the oscillations
with a damped sine-function of the same form as equation~(\ref{eqoscft}). The fitted 
oscillations and the background trends are shown in Figures~\ref{fgtnt}(a) and (b).
The detrended oscillations with the fitted parameters are shown in 
Figures~\ref{fgtnt}(c) and (d). We find that the temperature and density oscillations 
have similar periods and they are nearly in phase. The initial phase shift is 
$\Delta\phi=6^{\circ}\pm23^{\circ}$. The phase shift measured using the cross correlation
from the relative perturbations to the trend is about 12$^{\circ}$.

\begin{figure*}
\epsscale{1.0}
\plotone{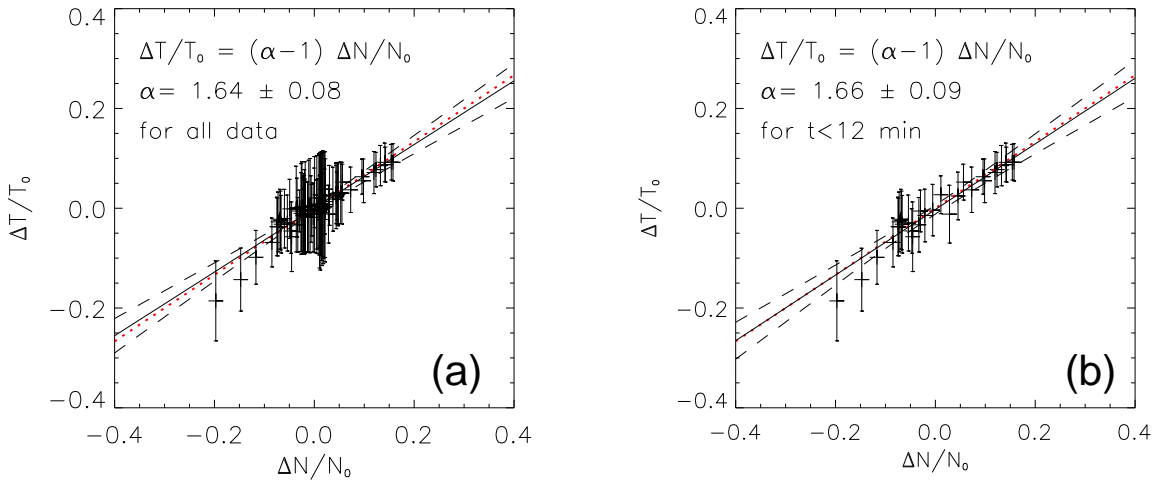}
\caption{ \label{fggma}
Measurements of the polytropic index $\alpha$. (a) The scatter plot of electron density 
and temperature perturbations (pluses), with the best-fitting line (solid) and 
the line of $\gamma$=5/3 (dotted). The dashed lines indicate the $\pm1\sigma$ fitting error.
(b) Same as (a) but for the data within 12 min after 12:46:20 UT.}
\end{figure*}

Under the polytropic assumption $p\sim n^{\alpha}$ with a polytropic index $\alpha$, 
it can be derived from linearized ideal MHD theory that \citep[e.g.][] {van11}: 
\begin{equation}
\frac{T^{'}}{T_0}=(\alpha-1)\frac{n^{'}}{n_0}, \label{eqfgm}
\end{equation}
where $p$, $n$, $T$ are the gas pressure, electron number density, and temperature, respectively. 
A superscript ($^{'}$) indicates perturbed quantities and a subscript ($_0$) stands for equilibrium
quantities. The process is adiabatic only if $\alpha$=$\gamma$=5/3.
The polytropic index can be measured using the linear relationship between the observables 
$T^{'}/T_0$ and $n^{'}/n_0$. Note that the equation~(\ref{eqfgm}) does not generally hold true 
in the nonideal MHD case. For example, thermal conduction can lead a large
phase shift between $T^{'}$ and $n^{'}$ (see the discussion in Section~\ref{sctdis}). 
In this case, the polytropic index should be measured either using equation~(\ref{eqfgm}) 
for the data after removing the phase shift, or based on the wave amplitude measurements as
\begin{equation}
\alpha=\frac{A_T(t)}{A_n(t)}+1,
\label{eqtgm}
\end {equation}
where $A_T(t)$ and $A_n(t)$ are the decaying temperature and density amplitudes normalized 
to the corresponding trend. Here we chose equation~(\ref{eqfgm}) to measure 
the polytropic index because the observed temperature and density perturbations are 
nearly in phase and the advantage compared to using equation~(\ref{eqtgm})
is that the assumption of the oscillations as a damped sine-function is not required. 
We obtained $\alpha=1.64\pm0.08$ using the linear least-squares fitting to the scatter plot 
of the data ($T^{'}/T_0(t)$ and $N^{'}/N_0(t)$) over the wave lifetime of 12:46--13:20 UT 
(Figure~\ref{fggma}(a)), and obtained $\alpha=1.66\pm0.09$ by fitting only the data within 
the first oscillation period (Figure~\ref{fggma}(b)). We find that the value of $\alpha$
agrees well with the adiabatic index of 5/3 for fully ionized coronal plasmas.
Note that the assumptions of the line-of-sight column depth and the filling factor used in 
determination of the loop density from EM have little effect on the accuracy of the 
measurement because $\alpha$ depends only on the relative perturbations of 
the density ($n^{'}/n_0$). 

\section{Discussion and Conclusions}
\label{sctdis}
We have studied a non-eruptive flare event (not associated with a CME) occurring 
near one footpoint of a large coronal loop observed by SDO/AIA. The flare impulsively heats 
the loop to above 10 MK and also generates the longitudinal waves in the hot loop. 
Despite of difficulties in distinguishing propagating and standing waves based on only 
intensity information, we argue that the observed wave is most likely of 
a fundamental standing slow mode. This is suggested by the signature that the temperature and 
density oscillations match well a harmonic wave with the frequency close to that for the 
fundamental mode. In contrast, a reflecting single wave pulse does not follow 
a sine-function in temporal variation and its time distance plot exhibits a `zigzag' 
pattern \citep{dem03, tar08, kum13}. 
In addition, the nearly inphase intensity oscillations shown along each leg 
after $\sim$12:50 UT suggest that the standing wave appears to form within one wave period 
after the flare, consistent with the case of standing slow waves observed by 
SOHO/SUMER \citep{wan03a, wan05}. The quick setup of the standing wave could be related to
meeting of the two oppositely-propagating brightenings (marked with arrows in 
Figure~\ref{fgosc}(b)) in the initial phase \citep{sel09}, however, a detailed
study on the wave excitation is beyond the scope of this Letter.
 
Here, we focus our analyses on the plasma thermal and wave properties of the hot loop, and
have measured the polytropic index $\alpha$ from the temperature and
electron density perturbations. We find that the value of $\alpha$ 
is close to 5/3 within a small uncertainty particularlly in the first oscillation period. 
This is striking because it implies that the energy equation in the MHD equations can
be well represented with an adiabatic form, or the non-ideal effects 
such as energy gain and losses on the timescale of oscillations 
are negligible in the hot plasma of $T\ga$9 MK. 
The reasons are detailed in the following based on the linearized 1D hydrodynamic
equations including the effects of gravity, heat conduction, compressive viscosity, and
optically thin radiative losses and gains \citep[e.g.,][]{dem03, sig07}. Note that 
both the measurements of $\alpha$ and the theoretical interpretations below 
are independent on whether the observed slow wave is of propagating or standing mode
because they are essentially identical in physical property.

(1) We know that for an isothermal loop in hydrostatic equilibrium, the pressure 
scale height $H=50T_6$ Mm. For the hot loop at $T\approx$9 MK, 
we have $H/h\approx$9 where $h$=50 Mm, the estimated loop height. 
Thus the stratification effect here can be neglected. 

(2) We estimate the effect of radiative loss
from the ratio of the oscillation period to the radiation time scale $r=P/\tau_{\rm rad}$. Here
$\tau_{\rm rad}=3450n_9^{-1}T_6^{3/2}$ s, where $n_9=n/[10^9 {\rm cm}^{-3}$] \citep{sun13}. 
For the measured parameters $T$=8.7 MK, $n=2.6\times10^9$ cm$^{-3}$ and $P$=12 min, 
we obtain $\tau_{\rm rad}\approx$570 min and $r\approx$0.02. Thus radiative cooling is negligible 
on the oscillation period timescale. 

(3) It is known that when thermal conduction dominates in the energy equation, it introduces 
a phase shift ($\Delta{\phi}$) between the density and temperature perturbations 
\citep{owe09, van11}. The linear approximation gives the following relations
\begin{eqnarray}
{\rm tan}\Delta{\phi}& = &\frac{\pi(\gamma-1)\kappa_\|}{k_Bc_s^2Pn}, \label{eqphi} \\
\frac{A_T}{A_n}& = &(\gamma-1){\rm cos}\Delta\phi ~~~[=\, \alpha-1], \label{eqgmp}
\end{eqnarray}
where $\gamma$=5/3, $k_B$ the Boltzmann constant, 
$\kappa_\|=7.8\times10^{-7}T^{5/2}$ ergs cm$^{-1}$s$^{-1}$K$^{-1}$ the thermal 
conductivity parallel to the magnetic field \citep{spi62}. Note that the above equations
hold true no matter whether thermal conduction or compressive viscosity dominates in the wave
damping (see item (4)). 1D MHD simulations have shown the presence of a large phase shift 
between $T^{'}$ and $n^{'}$ in hot loops due to thermal conduction \citep[e.g.][]{ofm02, sig07}.
 
However, we find that the temperature and density oscillations are nearly in phase. 
This result suggests that the thermal conduction may be much weaker than the classical 
theory predicted in the hot flaring loop.  With the loop parameters used in item (2) 
we obtain the theory-predicted $\Delta{\phi}=35^{\circ}$ using equation (\ref{eqphi}). 
The comparison of the observed phase shift ($\sim12^{\circ}$) with the predicted suggests 
a reduction of (parallel) thermal conductivity by a factor of three. Moreover,
from the measured polytropic index of $\alpha$=1.64 we derive the phase shift to be 
$\sim16^{\circ}$ using equation (\ref{eqgmp}), which is consistent with the
directly measured value. Thus both the phase shift and polytropic index measurements 
support the conduction suppression in this event. In addition, the results of $\alpha$=1.66
that is even closer to $\gamma$=5/3 and a better in-phase relationship between 
$T^{'}$ and $n^{'}$ within the first oscillation period suggest that thermal conduction
tends to be more suppressed when the plasma is hotter ($T\ga$9 MK).

In addition, we notice that for hot loop oscillations observed by SUMER and SDO/AIA 
\citep{wan03a, wan07, kum13, kum15}, the measured phase speeds are close to the adiabatic sound 
speed within uncertainties, also suggesting $\alpha\sim$5/3. This is in contrast to the
theoretical prediction from linear MHD simulations by \citet{dem03}. They showed that when
thermal conduction is dominant due to the high temperatures, the perturbations propagate
largely undamped, at the slower, isothemal sound speed, which implies $\alpha\sim$1.

(4) In using equation~(\ref{eqfgm}) to measure the polytropic index, 
we have assumed that the loop temperature (or sound speed) varies gradually with time 
relative to the wave period (known as the WKB approximation). To check its validity, 
we quantify the ``gradual variation'' as the requirement that changes in wave period must be small 
over a single period. The condition can be expressed as 
$|dP/dt|\ll1$. The fitting to the evolution of the loop temperature (Figure~\ref{fgtnt}(a)) 
gives the trend, $T(t)=T_0+at+bt^2$, where the initial temperature $T_0$=12.6 MK, 
the polynomial coefficients $a$=$-$0.38 MK~min$^{-1}$ and $b$=0.0064 MK~min$^{-2}$.
Considering $P(t)=2L/c_s(T(t))$, we obtain $|dP/dt|=P_0T_0^{1/2}|0.5a+bt|T(t)^{-3/2}$, 
where $P_0=2L/c_s(T_0)$=10.5 min. The calculation of its maximum value $|dP/dt|_{\rm max}$=0.16
indicates that the WKB approximation is valid in this case. This justifies the removal 
of the trend from the oscillations in our analysis. 

(5) As the viscous heating belongs to the second-order term in the linearized energy equation, 
its effect on wave damping can be neglected. Thus the only dominant damping 
mechanism left is the momentum loss by the viscous forces. From the dispersion relation 
derived from the velocity wave equation when only compressive viscosity is present
\citep[see Eq.(33) in][]{sig07}, we can obtain the coefficient of compressive viscosity
in terms of the observables by
\begin{equation}
 \eta_s=\frac{3\gamma p_0\tau}{8\pi^2(\tau/P)^2+2}
  = \frac{3\gamma k_Bn_0T_0\tau}{4\pi^2(\tau/P)^2+1}, \label{eqvis}
\end{equation}
where $\gamma$=5/3 and $p_0=2n_0k_BT_0$. By taking $P=12.0\pm0.6$ min and 
$\tau=9.2\pm2.1$ min (the mean for 
temperature and density oscillations) with $T_0=8.7\pm1.7$ MK and 
$n_0=(2.6\pm0.2)\times10^9$ cm$^{-3}$, we obtain $\eta_s=356\pm195$ g cm$^{-1}$ s$^{-1}$. 
In comparison, we calculate the classical {\it Braginskii} compressive viscosity coefficient using
$\eta_0=0.1 T_6^{5/2}$ g cm$^{-1}$ s$^{-1}$ \citep[e.g.,][]{ofm02}, and obtain
$\eta_0\approx23$ g cm$^{-1}$ s$^{-1}$, thus, $\eta_s/\eta_0=15$. This implies that
to interpret the wave damping timescale by the compressive viscosity alone the classical viscosity
coefficient needs to be enhanced by a factor of 15, which may be regarded as an
upper limit considering the additional effects such as weak nonlinearity, thermal
conduction and stratification.

In summary, we have found quantitative evidence of thermal conduction suppression in a hot
 flare loop by coronal seismology of the slow-mode waves. This result suggests that the flare
loop should cool much slower than expected from the classical {\it Spitzer} conductive cooling.
Our studied event is indeed of such long-duration events (LDEs) which are the flares with 
a slower-than-expected decay rate in soft X-ray and EUV radiation 
\citep[e.g.,][]{for96, tak00, qiu12}.
To explain the LDEs and flare loop-top sources, some previous studies have suggested the
mechanism of continuous heating \citep{war06, liu13, sun13} or conduction suppression 
\citep{mct93, jia06, li12}. Our study confirms the effect of the latter mechanism in a more 
direct way. Laboratory experiments and numerical studies showed that the actual
conductivity is smaller (by at least a factor of two) than that given by Spitzer when 
$l\la30\lambda$ where $l$ is the temperature gradient scale length and $\lambda$ the mean free
path of thermal electrons \citep{bel81, luc83}. We estimate $l/\lambda\sim30$ in this case,
suggesting that the nonlocal conduction may account for the observed conduction 
suppression \citep{mat82, ros86}. By studying the evolution of flare loop-top sources, 
\citet{jia06} suggested that plasma waves or turbulence may play an important role 
in suppressing the conduction during the decay phase of flares. The mechanism is similar
to that used for interpreting the significant reduction of thermal conductivity 
in galaxy-cluster cooling flows by a tangled magnetic field \citep{cha98}. 
Finally, we conclude that the result of conduction suppression may also shed 
light on the coronal heating problem \citep[see the review by][]{kli15}, because weak 
thermal conductivity implies smaller conductive losses, and an extended lifetime of 
individual nanoflares, increasing the average coronal temperature for the same heating rate.

The work of T. W. was supported by NASA grants NNX12AB34G and the NASA Cooperative Agreement
NNG11PL10A to CUA. L. O. and E. P. acknowledge to support from the NASA grant NNX12AB34G.
SDO is a mission for NASA’s Living With a Star (LWS) program.

\end{document}